\documentclass[a4paper,12pt]{article}

\usepackage{cmap}					
\usepackage{mathtext} 				
\usepackage[T2A]{fontenc}			
\usepackage[utf8]{inputenc}			
\usepackage[english]{babel}	
\usepackage{indentfirst}
\renewcommand{\epsilon}{\ensuremath{\varepsilon}}
\renewcommand{\phi}{\ensuremath{\varphi}}
\renewcommand{\kappa}{\ensuremath{\varkappa}}

\usepackage{amsmath,amsfonts,amssymb,amsthm,mathtools} 
\usepackage{icomma} 

\usepackage{graphicx}  
\graphicspath{{images/}{images2/}}  
\usepackage{wrapfig} 

\usepackage{array,tabularx,tabulary,booktabs} 
\usepackage{longtable}  
\usepackage{multirow} 

\theoremstyle{plain} 

\theoremstyle{definition} 

\theoremstyle{remark} 

\usepackage{etoolbox} 

\usepackage{extsizes} 
\usepackage{geometry} 
\usepackage{setspace} 

\usepackage{lastpage} 

\usepackage{soul} 

\usepackage{hyperref}
\usepackage[usenames,dvipsnames,svgnames,table,rgb]{xcolor}
\hypersetup{				
    unicode=true,           
    pdftitle={Заголовок},   
    pdfauthor={Автор},      
    pdfsubject={Тема},      
    pdfcreator={Создатель}, 
    pdfproducer={Производитель}, 
    pdfkeywords={keyword1} {key2} {key3}, 
    colorlinks=true,       	
    linkcolor=black,          
    citecolor=black,        
    filecolor=black,      
    urlcolor=black           
}

\usepackage{csquotes} 

\usepackage{multicol} 

\usepackage{tikz} 
\usepackage{pgfplots}
\usepackage{pgfplotstable}
\usepackage{float}
\usepackage{cite}
\usepackage{subcaption}
\title{Low-impedance stripline kicker for the transverse instability suppression system of the synchrotron radiation facility ``SKIF'' light source}

\author{ M.\,A.~Baistrukov$^{a,b}$\footnote{E-mail: M.A.Baistrukov@inp.nsk.su}, E.\,A.~Bekhtenev$^{a}$, A.\,A.~Krasnov$^{a,b}$,\\ D.\,A.~Nikiforov$^{a,b}$, P.\,A.~Piminov$^{a,b}$}

\date{$^a$\small{Budker Institute of Nuclear Physics SB RAS, Russia, Novosibirsk, 630090} \\  
$^b$\small{Synchrotron Radiation Facility --- Siberian Circular Photon Source ``SKIF'', Russia, Koltsovo, 630090} \\
}

\begin{document}

\maketitle

\begin{abstract}
Construction of the new fourth generation synchrotron radiation facility (SRF) ``SKIF'' near Novosibirsk is nearing completion.
``SKIF'' uniqueness is ultra-low 75~pm natural emittance at 400~mA beam current and 3~GeV beam energy.

Collective effects (beam interaction with the vacuum chamber) are the main obstacles of achieving such emittance and beam current.
To suppress these effects the storage ring will be equipped with a bunch-by-bunch feedback system.
Important components of the feedback system are a stripline (fast) kicker and a beam position monitor.

The paper describes design of the low-impedance stripline kicker for SRF ``SKIF'' storage ring, optimized for the manufacturing and for the uniformity of the imparted transverse kick.
The kicker qualities such as S-parameters, imparted transverse momentum, transfer and beam impedances are presented.
The chosen length of the stripline kicker allows its use as a beam position monitor.

\end{abstract}


\section{Introduction}
Construction of the new fourth generation synchrotron radiation facility (SRF) ``SKIF'' near Novosibirsk is nearing completion.
A key feature of the source is ultra-low 75~pm natural emittance at 400~mA beam current and 3~GeV beam energy \cite{SKIF_PRAB,SKIF_Crystal,SKIF_dynamic}.
Despite optimization of the vacuum chamber impedance, both multibunch and single-bunch instabilities occur in fourth-generation SR sources at certain beam currents.
These instabilities can lead to either an increase in the effective beam emittance or beam loss.
This, in turn, can limit operation in certain filling pattern modes.
Therefore, ``SKIF'' storage ring was designed with a bunch-by-bunch feedback system, allowing for partial suppression of multibunch instabilities.
A bunch-by-bunch feedback system also allows for partial suppression of strong single-bunch instabilities, such as TMCI, and to increase the threshold current \cite{feedbacks_1}.

Important components of the bunch-by-bunch feedback system are the stripline kicker and the beam position monitor.
To simplify the development and manufacturing process we decided to use the same design for the beam position monitor as for the stripline kicker.
To select the design of the stripline kicker, we relied on MAX~IV experience.
Initially, four-strip kickers with straight strip ends were developed for this facility \cite{kicker_M4_v1}.
However, during the operation of these kickers, the MAX~IV team concluded that it was better to use two-strip kickers with conical tapers at the strip ends \cite{kicker_M4_v2}.
Conical tapers at the ends of the strips ensured better matching of the vacuum feedthroughs to the strips and reduced the impedance of the stripline kicker.
We decided to develop a similar two-strip kicker for the transverse instability suppression system of SRF ``SKIF'' storage ring.

During the development of the stripline kicker model, we worked closely with engineers from the experimental production facility that will manufacture these stripline kickers.
As a result, the design of the stripline kicker was modified and adapted to the production technology.
These changes had little impact on the final characteristics of the stripline kicker, but should allow for faster and more labor-efficient production.

Section~\ref{Sec:Construct_features} describes the design features with a description of the manufacturing process, rationale for technical solutions and a description of the stripline kicker geometry.
Section~\ref{Sec:EM_field} clarifies the process of matching vacuum feedthroughs with striplines and describes briefly electromagnetic fields.
Section~\ref{Sec:Homogeneity} is devoted to optimizing the uniformity of the stripline kicker with respect to the transverse momentum imparted to the beam particles.
Section~\ref{Sec:BPM} gives the characteristics of the stripline kicker as a beam position monitor (BPM).
Section~\ref{Sec:Beam_impedance} describes the beam impedance of the stripline kicker.

\section{Design Features}
\label{Sec:Construct_features}

The stripline kicker is designed to not restrict the vacuum chamber aperture and to have low impedance.
Input from manufacturing engineers enabled us to develop a repairable design adapted to the production technology.
Furthermore, this design has virtually no impact on the basic characteristics of the stripline kicker.

\subsection{Production of stripline kicker}

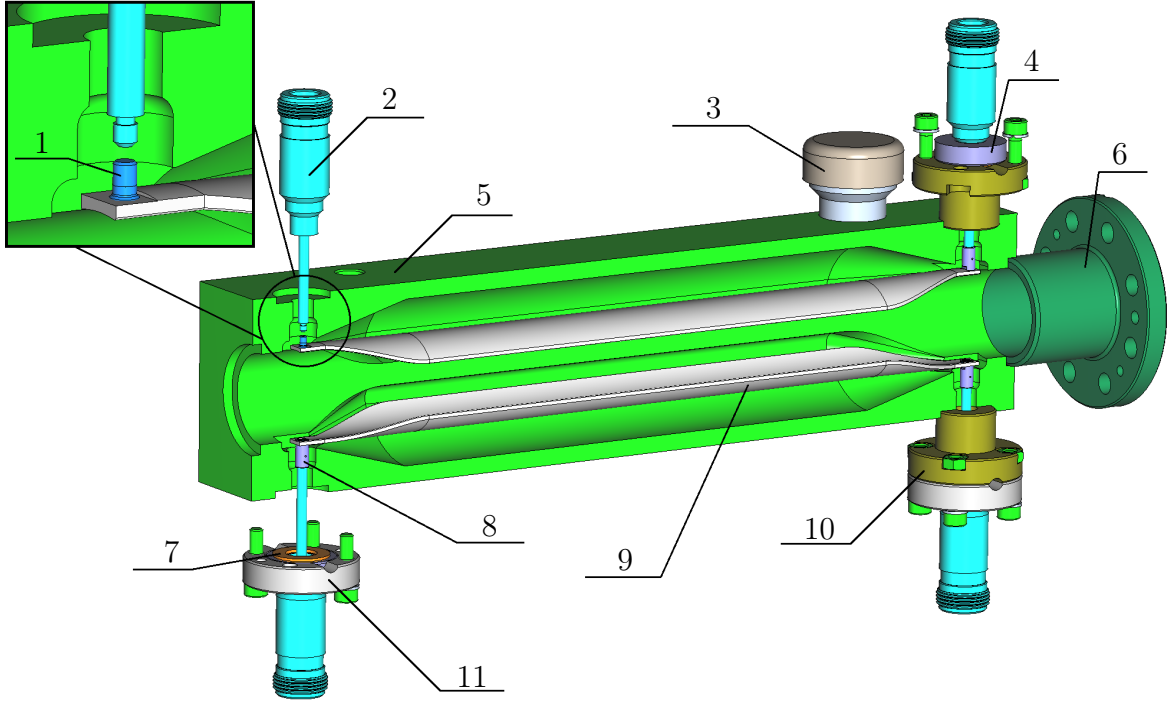
\begin{figure}[h]
    \centering
    \begin{picture}(1\textwidth ,0.638\textwidth)(0,0)
    \put(0,0){
    \includegraphics[width=1\textwidth]{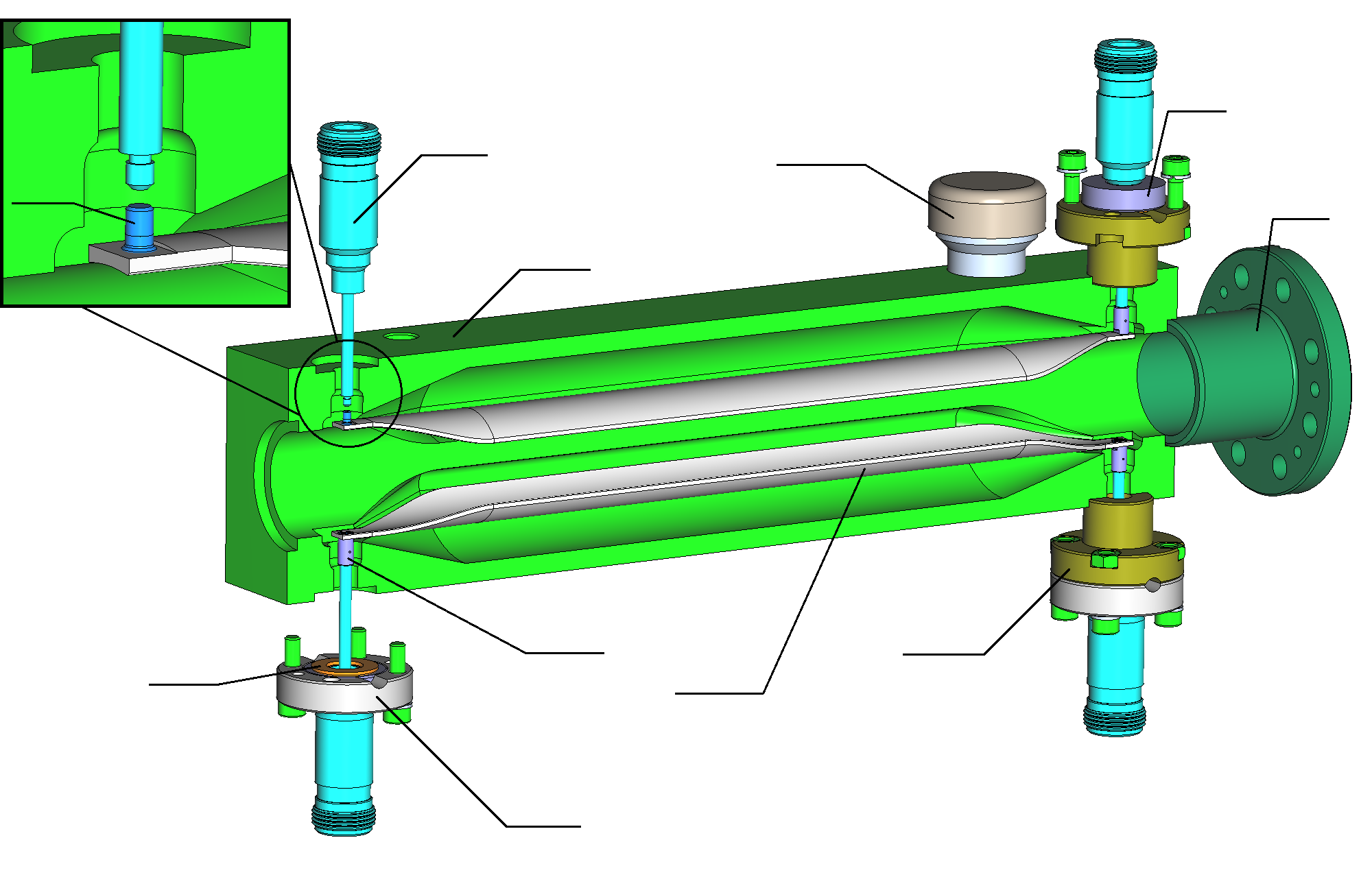}}
    \put(0.04\textwidth ,0.495\textwidth ){\makebox(0,0)[cb]{1}}
    \put(0.335\textwidth ,0.532\textwidth ){\makebox(0,0)[cb]{2}}
    \put(0.61\textwidth ,0.525\textwidth ){\makebox(0,0)[cb]{3}}
    \put(0.88\textwidth ,0.565\textwidth ){\makebox(0,0)[cb]{4}}
    \put(0.415\textwidth ,0.45\textwidth ){\makebox(0,0)[cb]{5}}
    \put(0.955\textwidth ,0.485\textwidth ){\makebox(0,0)[cb]{6}}
    \put(0.15\textwidth ,0.145\textwidth ){\makebox(0,0)[cb]{7}}
    \put(0.42\textwidth ,0.167\textwidth ){\makebox(0,0)[cb]{8}}
    \put(0.535\textwidth ,0.138\textwidth ){\makebox(0,0)[cb]{9}}
    \put(0.7\textwidth ,0.165\textwidth ){\makebox(0,0)[cb]{10}}
    \put(0.405\textwidth ,0.038\textwidth ){\makebox(0,0)[cb]{11}}

    \end{picture}
    \caption{General view of the stripline kicker. Some parts of the kicker are hidden to demonstrate the design. The main components of the stripline kicker are labeled with numbers.}
    \label{fig:Common_view}
\end{figure}

Figure~\ref{fig:Common_view} shows the general view of the stripline kicker with some details hidden to demonstrate the internal structure.
The main components of the stripline kicker are:
1\,--\,mounting screw, 2\,--\,vacuum feedthrough, 3\,--\,geodetic sign, 4\,--\,support flange, 5\,--\,half of the housing, 6\,--\,vacuum chamber flange, 7\,--\,vacuum gasket, 8\,--\,intermediate electrode, 9\,--\,strip, 10\,--\,input flange, 11\,--\,slip-on flange.

We propose the following manufacturing process.
First, the two halves of the housing are machined on a milling machine.
The vacuum cavity shape is selected so that the main part of the cavity is machined with a 4 mm radius ball end mill, while the conical narrowing and transition to the inlet are machined with a 2 mm radius ball end mill and finished with a 2 mm radius cylindrical end mill.
Then, the halves are welded together, and holes for the vacuum inlets are drilled in the housing, grooves for the vacuum chamber flanges and the inlet flanges are bored, and holes for the geodetic signs are drilled.
Then, the vacuum chamber flanges and the input flanges are welded to the housing.

Vacuum feedthroughs are modified as follows.
A thread for the intermediate electrode is cut on the end of the electrode.
Then, a slip-on flange is placed on the feedthrough, after which a support flange is welded to the feedthrough.

The strips are manufactured as follows.
A cylindrical tube is machined on a lathe to ensure that the longitudinal profile of the tube's outer diameter matches the longitudinal profile of the strip's thickness.
Next, two strips are cut from the resulting blank using electrical discharge machining.
The sharp edges are rounded, and holes and recesses for the mounting screw and a chamfer for the intermediate electrode are made in the strip.
The intermediate electrodes are then attached to the strip using a mounting screw.

Final assembly is performed as follows.
The strip is inserted into the housing through the vacuum chamber flanges and secured in the assembly position with a tool.
Next, vacuum gaskets are placed on the input flanges, and the modified vacuum feedthroughs are inserted into them.
The vacuum feedthroughs are screwed to the intermediate electrodes of the strip.
The tooling is then removed, and the vacuum feedthroughs are moved to the operating position.
The union flange is tightly secured to the input flange with bolts and nuts.
The procedure is then repeated for the second strip.
Finally, geodetic signs are attached to the housing.

\subsection{Justification of technical solutions}
This design used several technical solutions that may require clarification.
Here is a list of the main technical solutions and the rationale for their use:
\begin{enumerate}
\item Using ball-end mills to cut the main cavity and, accordingly, round the edges of the main vacuum cavity. Rounding the edges of the vacuum cavity does not lead to any significant differences in the field distribution, as simulations have shown, and significantly simplifies the manufacturing process.
\item Dividing the housing into two halves vertically, longitudinally. This cuts out two halves of the vacuum cavity in each half. Manufacture a solid housing without dividing it in half is not possible on our production equipment. The simplest solution for manufacturing several devices is to split the housing into two halves and use a milling machine. Splitting horizontally, however, makes it impossible to manufacture the housing halves for us.
\item Machining individual input flanges on a lathe and welding them to the housing. This approach ensures the required roughness level of the vacuum seal. It also allows for repeated assembly and disassembly of the stripline kicker. The option of attaching the vacuum feedthrough directly to the housing was discarded during discussion. This was due to the impossibility of achieving the required level of roughness in the area of the vacuum seal, as well as the risk of damaging the housing threads during disassembly of the stripline kicker.
\item Using electrical discharge machining (EDM) to produce strips from a cylindrical vacuum chamber is the simplest and most natural approach. The alternative is stamping or milling the strips. However, stamping requires extensive preparation and is impractical for producing only a few devices. A strip, on the other hand, is a complex component for a milling machine and also requires extensive preparation.
\item Using an intermediate electrode to connect the strip to the vacuum feedthrough electrode. The reason for this approach is the assembly process. It is not possible to thread the vacuum feedthrough electrode into the strip itself. Attaching the strip directly to the electrode with a screw is extremely difficult, as the screw diameter is only slightly smaller than the vacuum feedthrough electrode diameter. Furthermore, holding the screw steady through the vacuum feedthrough flange is difficult.
\end{enumerate}

\subsection{Stripline geometry}
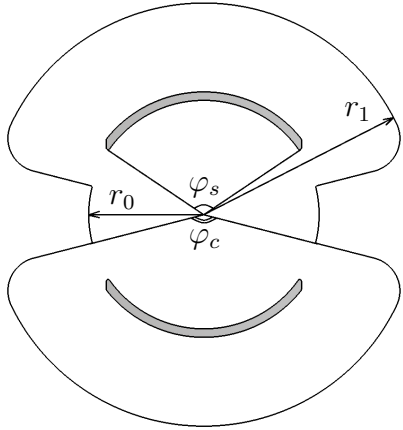
\begin{figure}[h]
    \centering
    \begin{picture}(0.4\textwidth ,0.4\textwidth)(0,0)
    \put(0,0){
    \includegraphics[width=0.4\textwidth]{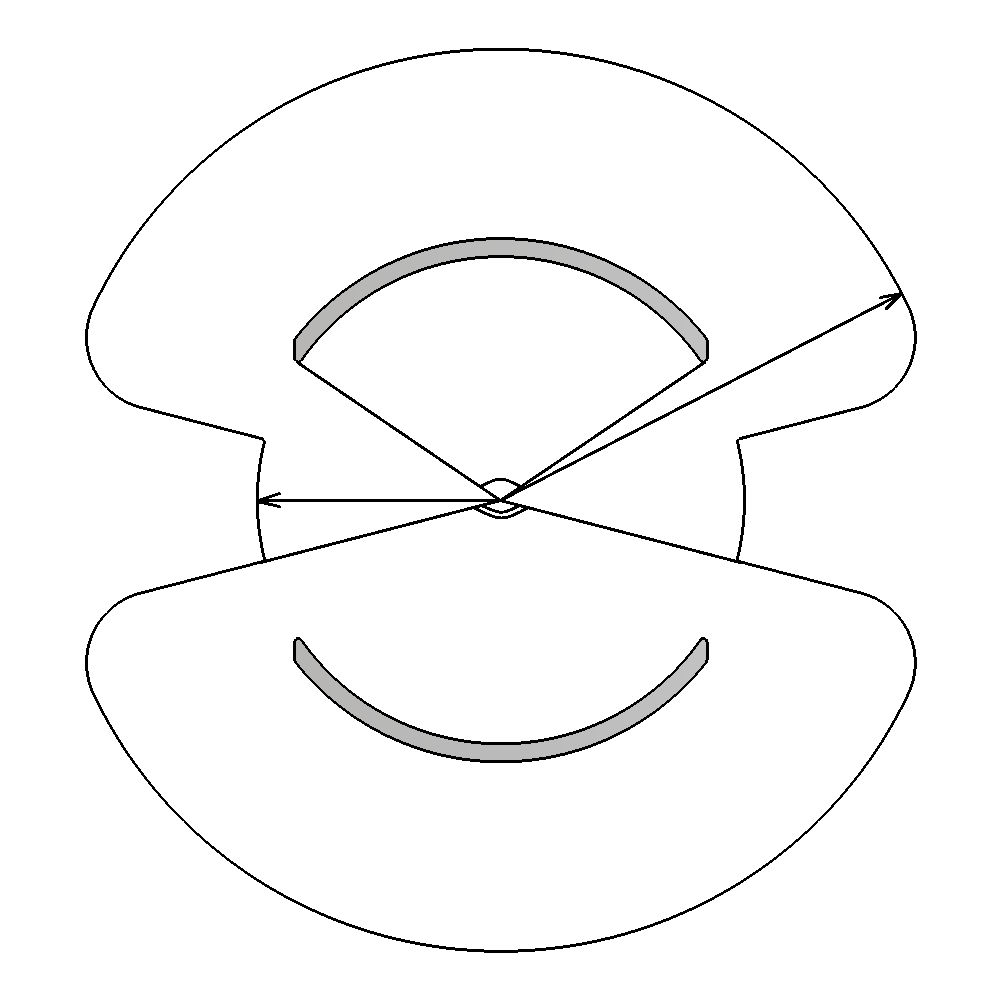}}
    \put(0.34\textwidth ,0.28\textwidth ){\makebox(0,0)[cb]{$r_1$}}
    \put(0.14\textwidth ,0.205\textwidth ){\makebox(0,0)[cb]{$r_0$}}
    \put(0.21\textwidth ,0.215\textwidth ){\makebox(0,0)[cb]{$\varphi_s$}}
    \put(0.21\textwidth ,0.17\textwidth ){\makebox(0,0)[cb]{$\varphi_c$}}
    
    \end{picture}
    \caption{Cross-section of the main stripline with principal dimensions.}
    \label{fig:2D_cross_section}
\end{figure}

Figure~\ref{fig:2D_cross_section} shows a cross-section of the main stripline with principal dimensions.
The angular size of the strip is $\varphi_s=115.6$~degrees.
The angular size of the cavity is $\varphi_c=151.1$~degrees.
The angular distance between the edge of the cavity and the strip is $\varphi_g=(\varphi_c-\varphi_s)/2=17.75$~degrees.
The internal radius of the vacuum chamber is $r_0=13.5$~mm.
The internal radius of the strip cavity is $r_1=25$~mm.
The edges of the cavity are rounded with a radius of 4~mm.
The sharp edges of the strip are rounded with a radius of 0.2~mm.

The width of the cavity at the narrow part is 9.21~mm, and the width of the strip is 5.4~mm.
The half aperture at the narrow part is 16.5~mm.
The length of the main part of the strip is 158~mm, the transition length is 30~mm, and the distance from input to input is 230~mm.
The thickness of the main part of the strip is 1~mm, increasing linearly to 1.3~mm at the transition.
At the attachment point of the intermediate electrode, a 0.3~mm deep chamfer is removed from the narrow part of the strip.
The radius of the intermediate electrode is 2~mm, and the length is 7~mm.
The length of the main part of the cavity is 160~mm, and the transition length is 30~mm.
The transition of the cavity is a conical narrowing with a linear decrease in the radius of rounding from 4~mm to 2~mm.

\section{Line matching and electromagnetic fields}
\label{Sec:EM_field}
\subsection{Impedance matching}
One of the problems that arises during the development of a stripline kicker is matching the striplines to the vacuum feedthrough.
The central part of the stripline kicker is essentially a stripline with two strips.
This stripline has two operating modes.
For distinctness, we assume that these are even (in-phase) and odd (antiphase) operating modes.
In the first case, the voltage across the strips is the same, while in the second, it is opposite in sign.
In these modes, the characteristic impedance is slightly different.

To minimize the reflected signal when operating a stripline kicker, it is necessary to match the entire line from the input to the output port.
The characteristic impedance of vacuum feedthroughs is 50~ohms.
Accordingly, the characteristic impedance of the narrow stripline and the main stripline must also be 50~ohms.
Furthermore, the shape of the transition from the narrow stripline to the main stripline must be selected to minimize reflection of the transmitted signal.

Characteristic impedance of a stripline $Z_s=\sqrt{L/C}$.
The phase velocity \mbox{$v_p=1/\sqrt{L C}$} coincides with the speed of light $c_0$ for a vacuum stripline.
Here $L$ and $C$ are the inductance and capacitance per unit length.
The traditional approach to finding $Z_s$ is to solve the electrostatic problem and find $C$.
Knowing $C$, we can calculate $Z_s=1/( C c_0)$.
For simple geometry, where the angular size of the strip is constant throughout its thickness, and the angular size of the cavity is constant throughout its depth, an analytical solution for the electrostatic potential can be found.
The solution is an infinite series.
The solution accuracy depends on the number of considered terms in the expansion of the solution for the region along the radius between $r_0$ and $r_0+h$, where $h$ is the thickness of the strip.

However, due to the strip design and the rounded cavity edges, in our case this approach gives only a rough estimate.
For this reason, the Time Domain Solver for High Frequency problems from the CST Studio suite was used to calculate the characteristic impedance of both the narrow and main strips.
The stripline characteristic impedance $Z_s$ was calculated using the linear impedance $Z_l$, which was calculated during the simulation in CST Studio suite.
Since we do not plan to use the stripline kickers in the even mode, stripline matching was performed in the odd mode.
For this, the appropriate symmetry conditions were specified in CST Studio.
For a vertical kicker, this means the absence of tangential electric fields in the longitudinal horizontal plane and the absence of tangential magnetic fields in the vertical longitudinal plane.
In this mode, $Z_s=Z_l/2$.

By varying the parameters for the main and narrow stripline we found the numerical dependence of $Z_s$ on the parameters.
Based on this dependence, we constructed surfaces in the parameter space where $Z_s=50$~Ohms.
Next, we selected starting points on this surface and constructed a transition between the narrow and main sections of the stripline.
The transition was defined by several parameters.
To optimize the matching, we used time-domain reflectometer (TDR) analysis, which reconstructs the line impedance from the reflected signal.
Finally, we found the transition parameter values that ensured the best matching.
After completing the stripline kicker model and optimizing it for uniformity of the transmitted transverse pulse, the transition was again optimized for impedance matching.

\begin{figure}[h]
    \centering
    \includegraphics{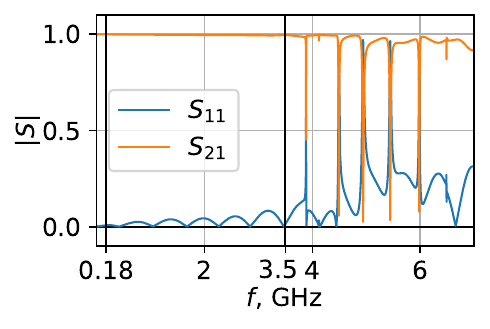}
    \caption{Calculation of the $S$ parameter from input to input in antiphase mode. $S_{11}$ is the reflected wave, $S_{21}$ is the transmitted wave.}
    \label{fig:s_param}
\end{figure}

The results of modeling the S reflection and transmission parameters in odd mode operation for the stripline kicker are shown in Figure~\ref{fig:s_param}.
At a frequency of $f=180$~MHz, the negative feedback amplifier bandwidth, reflection from the ports can be neglected.
Furthermore, if necessary, operation can be carried out at frequencies up to $3.5$~GHz.

\subsection{Electromagnetic fields}
\begin{figure}[h]
    \centering
    \begin{picture}(1\textwidth ,0.4\textwidth)(0,0)
    \put(0,0){
    \includegraphics[height=0.41\textwidth]{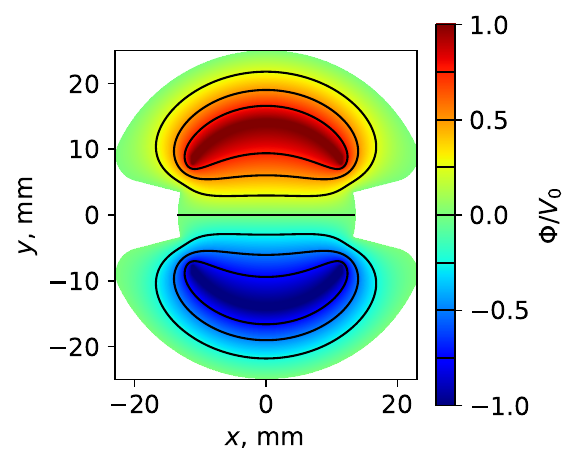}}
    \put(0.51\textwidth,0){
    \includegraphics[height=0.41\textwidth]{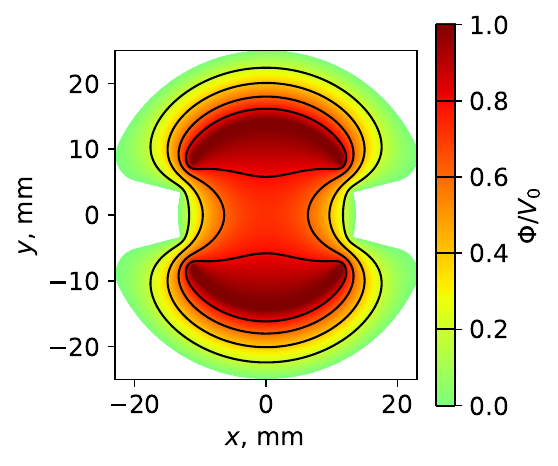}}
    
    \put(0.06\textwidth, 0.36\textwidth){\makebox(0,0)[lb]{(a)}}
    \put(0.56\textwidth, 0.36\textwidth){\makebox(0,0)[lb]{(b)}}
    \end{picture}
    \caption{Normalized electrostatic potential in odd (a) and even (b) mode.}
    \label{fig:2D_E_fields}
\end{figure}

The figure~\ref{fig:2D_E_fields} shows the electrostatic potential $\Phi$ normalized to the strip voltage $V_0$ in the odd~(a) and even~(b) regimes for a long stripline with geometry corresponding to the main stripline of the kicker.
The electric field $\vec{E}=-\nabla \Phi$.
For a TEM wave propagating in the direction of $\vec{n}$, the magnetic induction $\vec{B}=\vec{n}\times \vec{E}/c_0$.

Similar to the works~\cite{kicker_M4_v1,kicker_M4_v2}, we introduced the longitudinal $g_\parallel$ and transverse $g_\perp$ geometric factors\cite{kicker_theory}:
\begin{equation}
    \label{eq:g_long}
    g_\parallel=\frac{\Phi(\rho=0)}{V_0}=0.72,
\end{equation}
\begin{equation}
    \label{eq:g_perp}
    g_\perp=\frac{r_0 E_y(\rho=0)}{V_0}=1.14,
\end{equation}
where $\Phi$ is the electric potential in the even mode, $E_y$ is the vertical electric field in the odd mode, $\rho$ is the distance from the center in the polar coordinate system, $V_0$ is the voltage on the strip.
It is also interesting to compare the geometric factors with the analytical expressions of $g_{\parallel,s}$ and $g_{\perp,s}$ for a simplified model of a similar strip kicker~\cite{kicker_M4_v2}.
\begin{equation}
    \label{eq:g_long_s}
    g_{\parallel,s}=\frac{\varphi_s+\varphi_g}{\pi}=0.74,
\end{equation}
\begin{equation}
    \label{eq:g_perp_s}
    g_{\perp,s}=\frac{8}{\varphi_g \pi}\sin\left( \frac{\varphi_g}{2} \right) \sin\left( \frac{\varphi_s+\varphi_g}{2} \right)=1.16.
\end{equation}
The discrepancy with the simplified model begins already in the second digit of the fractional part, which reinforces the weak applicability of the analytical solution of the simplified model in our case.

\section{Kicker uniformity optimizing}
\label{Sec:Homogeneity}
When designing a stripline kicker, it is necessary to select the appropriate angular size for the strips.
The question arises: what angular size we should consider appropriate?
As a first approximation, the uniformity of the electric field near the kicker axis can be introduced as a criterion.
If the edges of the kicker strips can be neglected, then this approach is correct.
However, in our case, the narrow part of the strip, together with the transition from the narrow to the main part, makes a significant contribution to the transverse momentum imparted to the beam particles.
Therefore, the stripline kicker should be optimized for the uniformity of the transverse momentum imparted to the beam particles during the kicker's passage.

To find an approximation for the dependence of the momentum imparted to the beam particles near the kicker axis, we carried out some analytical calculations.
We assume that the velocity of the particle $\vec{v}$ is parallel to the kicker axis and is constant.
Then the Lorentz force is
\begin{equation}
    \label{eq:F_Lorentz}
    \vec{F}=q\left(\vec{E}+\vec{v}\times\vec{B}\right),
\end{equation}
where $q$ is the beam particle charge, $\vec{E}$ is the electric field, and $\vec{B}$ is the magnetic induction.
Since $\Box\vec{E}=0$, $\Box\vec{B}=0$, and $\vec{v}=\mathrm{const}$, it is also true that $\Box\vec{F}=0$.
Here $\Box$ is the d'Alembert operator.
\begin{equation}
    \label{eq:Dalamber}
    \Box=\Delta-\frac{\partial^2}{c_0^2 \partial t^2}=\Delta_2+\frac{\partial^2}{\partial z^2}-\frac{\partial^2}{c_0^2 \partial t^2},
\end{equation}
where $\Delta$ is the three-dimensional Laplace operator, and $\Delta_2$ is the two-dimensional Laplace operator.
We introduce a Cartesian coordinate system such that the stripline kicker axis is directed along the z-axis, the origin coincides with the center of the stripline kicker, and the electric field in the odd mode is directed along the y-axis.
We assume that the particle velocity is directed along the z-axis, and the longitudinal coordinate of the particle is $z=c_0 t$.
The momentum $\vec{P}$ that the particle receives when passing through the kicker is calculated as follows:
\begin{equation}
    \label{eq:P_vec}
    \vec{P}(x,y)=\int\limits_{-\infty}^{\infty}\vec{F}\left( x,y,z,t=\frac{z}{c_0} \right)\frac{\mathrm{d}z}{c_0}.
\end{equation}
We apply the two-dimensional Laplace operator to $\vec{P}(x,y)$:
\begin{equation}
    \label{eq:P_vec_laplace}
    \Delta_2\vec{P}(x,y)=\int\limits_{-\infty}^{\infty}\Delta_2\vec{F}\left( x,y,z,t=\frac{z}{c_0} \right)\frac{\mathrm{d}z}{c_0}=
    \int\limits_{-\infty}^{\infty}\left.\left( \frac{\partial^2}{c_0^2 \partial t^2}-\frac{\partial^2}{\partial z^2} \right)\vec{F}\left( x,y,z,t \right)\right|_{t={z}/{c_0}}\frac{\mathrm{d}z}{c_0}.
\end{equation}
Let's first consider the situation where the voltage on the kicker does not change between adjacent bunches.
This is the stationary case, $\partial/{\partial t}=0$ and \mbox{$\vec{F}(x,y,z,t)= \vec{F}_0(x,y,z)$}.
We substitute this condition into the right-hand side of \eqref{eq:P_vec_laplace} and integrate.
\begin{equation}
    \label{eq:P_vec_laplace_st}
    -\Delta_2\vec{P}(x,y)=
    \int\limits_{-\infty}^{\infty}\frac{\partial^2}{\partial z^2}\vec{F}_0\left( x,y,z \right)\frac{\mathrm{d}z}{c_0}=
    \frac{1}{c_0}\left.\frac{\partial}{\partial z}\vec{F}_0\left( x,y,z \right)\right|_{z\to -\infty}^{z\to \infty}=0.
\end{equation}
Here, equality to zero arises because the Lorentz force, together with its longitudinal derivative, tends to zero at an infinite distance from the stripline kicker.

Let's consider the case where a TEM wave travels along the stripline toward the beam.
Then the Lorentz force in the vacuum chamber can be approximately described by the following relationship:
\begin{equation}
    \label{eq:F_Lorentz_aprox}
    \vec{F}(x,y,z,t) = \vec{F}_0(x,y,z) A(z+c_0 t),
\end{equation}
where $A(z+c_0 t)$ is a multiplier describing the shape of the signal applied to the kicker.
Applying the two-dimensional Laplace operator to $\vec{P}(x,y)$, we receive:
\begin{multline}
\label{eq:P_vec_laplace_dinamic}
-\Delta_2\vec{P}(x,y)\approx\int\limits_{-\infty}^{\infty}\left.\left( \frac{\partial^2}{\partial z^2} - \frac{\partial^2}{c_0^2 \partial t^2} \right)\vec{F}_0\left( x,y,z \right)A(z+c_0 t)\right|_{t={z}/{c_0}}\frac{\mathrm{d}z}{c_0}= \\
=\int\limits_{-\infty}^{\infty}\left.\left(
\frac{\partial^2\vec{F}_0\left( x,y,z \right)}{\partial z^2}A(z+c_0 t) + 
2\frac{\partial\vec{F}_0\left( x,y,z \right)}{\partial z} A'(z+c_0 t)
 \right)\right|_{t={z}/{c_0}}\frac{\mathrm{d}z}{c_0} = \\
 =\int\limits_{-\infty}^{\infty}\frac{\partial}{\partial z}\left(\frac{\partial\vec{F}_0\left( x,y,z \right)}{\partial z} A(2z)\right)\frac{\mathrm{d}z}{c_0}=
 \frac{1}{c_0}\left.\left(\frac{\partial\vec{F}_0\left( x,y,z \right)}{\partial z} A(2z)\right)\right|_{z\to -\infty}^{z\to \infty}=0.
\end{multline}
Therefore, taking into account the symmetry of the stripline kicker, within our approximation, the transverse momentum imparted to the beam particles near the kicker axis can be expanded as follows:
\begin{equation}
    \label{eq:kicker_p_y}
    P_y= A_0 \left( 1 + A_1 \frac{x^2-y^2}{r_0^2} + A_2 \frac{x^4-6x^2 y^2+y^4}{r_0^4} \right),
\end{equation}
where $A_0$, $A_1$, $A_2$ are constant coefficients.
They depend on the signal shape, and $A_0$ also depends on the signal amplitude.
x and y are the horizontal and vertical displacements of the beam from the axis, respectively.

To optimize the uniformity of the stripline kicker, we should select the parameters such that $A_1=0$.
However, the coefficient $A_1$ depends on the shape of the signal supplied to the kicker, so this condition can only be satisfied for one operating mode.
Of all the operating modes, two extreme ones can be distinguished: the stationary mode and the mode of operation at maximum frequency.
We call the coefficients $A_i$ for the stationary mode $K_i$, and for the operating mode at maximum frequency we call them $D_i$.
We vary the angles $\varphi_s$ and $\varphi_c$ while maintaining the characteristic impedance so as to satisfy the condition $D_1=0$.
Then we slightly vary the transition parameters from the narrow strip to the main strip to improve matching at the new point.
One such cycle is sufficient to achieve a value of $D_1$ close to zero.
The results of numerical calculations of $K_i$ and $D_i$ for the stripline kicker model are given in table~\ref{tab:puls_char}.

\begin{table}[h]
    \caption{Calculated results of the transverse momentum parameters imparted to beam particles by a stripline kicker. The values of $K_0$ and $D_0$ are normalized to the peak power of the signal arriving at one port.}
    \label{tab:puls_char}
        \begin{center}
        \begin{small}
            \begin{tabular}{|c|c|c|c|}
                \hline
                Coefficient & Value & Coefficient & Value \\ \hline
                $K_0$ & $249\frac{\mathrm{eV}}{c_0\sqrt{\mathrm{W}}}$ & $D_0$ &  $223\frac{\mathrm{eV}}{c_0\sqrt{\mathrm{W}}}$ \\ \hline
                $K_1$ & $-0,059$ & $D_1$ &  $-0,0024$ \\ \hline
                $K_2$ & $-0,469$ & $D_2$  & $-0,507$ \\ \hline
                $K_0/p_{\parallel}$ & $82,9 \cdot 10^{-9} \frac{1}{\sqrt{W}}$ & $D_0/p_{\parallel}$ &  $74,5 \cdot 10^{-9} \frac{1}{\sqrt{W}}$ \\ \hline 
                $K_0\sqrt{N_{max}}/p_{\parallel}$ & $2,62\cdot 10^{-6}$ & $D_0\sqrt{N_{max}}/p_{\parallel}$ &  $2,35\cdot 10^{-6}$ \\ \hline 
            \end{tabular}
        \end{small}
        \end{center}
\end{table}

It can be seen that the coefficient $D_1$ is an order of magnitude smaller than the coefficient $K_1$, which corresponds to our optimization process.
It is also evident that when the amplifier operates at maximum power $N_{max}$, the transverse angle imparted to the beam particles in one passage through the stripline kicker lies in the range from $2,35$~\textmu rad to $2,62$~\textmu rad, depending on the kicker operating mode.

The transverse momentum imparted to the particles is often characterized by the transverse shunt impedance $R_\perp$\cite{kicker_M4_v1,kicker_M4_v2,kicker_theory}.
This quantity relates the total input RMS power $N_{rms}$ and the effective transverse voltage.
In our notation
\begin{equation}
    \label{eq:R_perp}
    R_\perp = \frac{A_0^2 c_0^2}{2 q^2 N_{rms}}.
\end{equation}
Strictly speaking, $R_\perp$ is a function of the frequency of the input signal, but in this paper we restrict ourselves to numerical values for extreme operating modes.
For the stationary operating mode $R_\perp=31$~kOhm, and for the operating mode at maximum frequency $R_\perp=24.8$~kOhm.

\begin{figure}[h]
    \centering
    \begin{picture}(\textwidth, 0.38\textwidth)(0,0)
    \put(-0.02\textwidth,0){
    \includegraphics[width=0.5\textwidth]{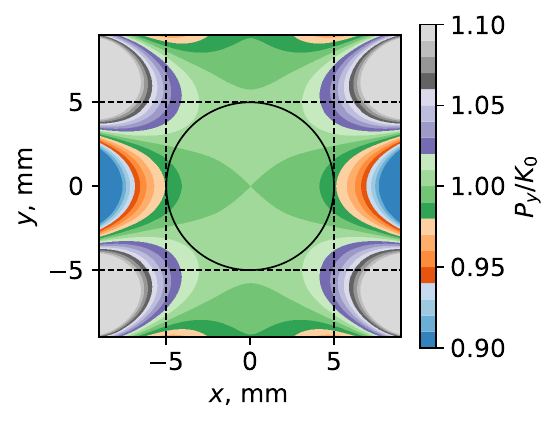}}
    
    \put(0.5\textwidth,0){
    \includegraphics[width=0.5\textwidth]{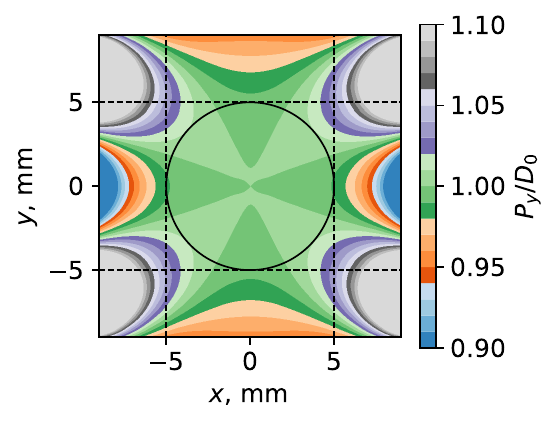}}
    
    \put(0,0.35\textwidth){\makebox(0,0)[lb]{(a)}}
    \put(0.52\textwidth, 0.35 \textwidth){\makebox(0,0)[lb]{(b)}}
    
    \end{picture}
    \caption{Uniformity of the vertical momentum imparted to the beam particles by the vertical stripline kicker in the stationary operating mode (a) and in the operating mode at the maximum frequency (b).}
    \label{fig:kicker_p_y}
\end{figure}

For the vertical stripline kicker model, numerical simulation was performed and $P_y(x,y)$ was calculated in two operating modes.
The figure~\ref{fig:kicker_p_y} shows color maps of the transverse pulse uniformity in the stationary operating mode (a) and in the operating mode at maximum frequency (b).
One color corresponds to a change of 1\%.
It can be seen that within a circle of radius 5~mm, the deviation of the transverse momentum imparted to the beam particles does not exceed 2\%.

\section{BPM characteristics}
\label{Sec:BPM}
The length of the strip is $l_0$.
Let's consider a simplified model of a stripline beam position monitor.
A passing charged particle induces a signal at the first port when it reaches the stripline region.
This signal then propagates along with the particle to the end of the stripline.
When the particle leaves the stripline region, it induces an opposite signal at the second port.
This signal compensates the first signal and propagates back along the stripline.
Thus, after $t=2 l_0/c_0$ from the initial signal, a signal with the opposite sign arrives at the first port.
If we denote the voltage on the first port as $u_1(t)$, and the voltage of the signal from the charged particle as $u_0(t)$, then $u_1(t)=u_0(t)-u_0(t-2 l_0/c_0)$.

In a cyclic accelerator, the greatest signal power usually falls on harmonics that are multiples of the main RF system frequency $f_{RF}$.
Accordingly, to obtain the greatest power at the first harmonic, it is necessary to satisfy the condition $2 f_{RF} l_0/c_0=1/2$.
This condition leads to the relation on $l_0$:
\begin{equation}
    \label{eq:l_0_simple}
    l_0 = \frac{\lambda_{RF}}{4},
\end{equation}
where $\lambda_{RF}=c_0/f_{RF}$ is the wavelength of the main RF system.
The length \eqref{eq:l_0_simple} is approximately 2 times shorter than the maximum length for an effective stripline kicker.
Therefore, it was the use of a stripline kicker as a beam position monitor that determined the length of the stripline kicker.
Due to the design features of our kicker, the situation is more complex, and a simplified model has limited applicability.
Therefore, the length of the strips was chosen so that the effective length of the strips $l_{\mathrm{eff}}$ was close to $\lambda_{RF}/4\approx 21$~cm.
The effective length of the vertical kicker is $l_{\mathrm{eff}}=c_0 P_y(0,0)/F_{0,y}(0,0,0)$, where $F_{0,y}(0,0,0)$ is the vertical Lorentz force acting on a particle at the center of the stripline kicker in the stationary mode.
In terms of meaning, $l_{\mathrm{eff}}$ is equivalent to the length of the stripline in the simplified stripline kicker model.

\begin{figure}[h]
    \centering
    \includegraphics[width=0.5\textwidth]{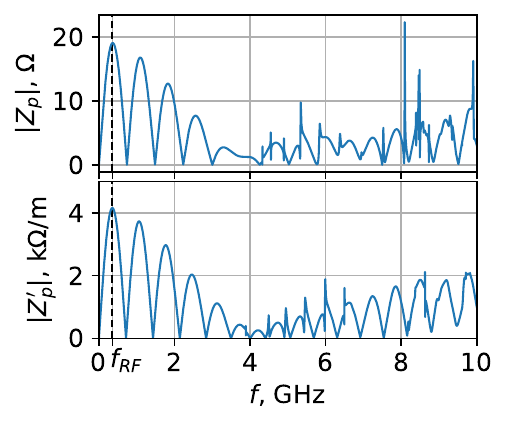}
    \caption{Longitudinal $Z_p$ and transverse $Z_p'$ transfer impedance of a stripline BPM. The vertical dash line corresponds to $f_{RF}$.}
    \label{fig:Transfer_impedance}
\end{figure}

The BPM is characterized by transfer impedance \cite{kicker_theory}.
The longitudinal transfer impedance $Z_p(\omega)=\hat{u}_1(\omega)/\hat{I}_b(\omega)$, where  $\hat{u}_1(\omega)$ is the Fourier transform of the voltage signal at the port, and $\hat{I}_b(\omega)$ is the Fourier transform of the bunch current inducing the signal at the port.
Obviously, when the bunch is shifted by $\Delta y>0$ toward the upper strip, the signals on the strips will differ.
To characterize the sensitivity of a stripline BPM to transverse displacement, a transverse transfer impedance is introduced
\begin{equation}
    \label{eq:Z_p_tr}
    Z_p'(\omega) = \frac{Z_p^+(\omega)-Z_p^-(\omega)}{\Delta y},
\end{equation}
where $Z_p^+(\omega)$ and $Z_p^-(\omega)$ are the longitudinal transfer impedances for a displaced bunch on the upper and lower strips, respectively.

We carried out simulations in CST studio to calculate the transfer impedance of the stripline kicker model.
The calculation results are presented in the figure~\ref{fig:Transfer_impedance}.
From the figure it can be seen that $Z_p$ and $Z_p'$ reach their maximum at $f_{RF}$ confirming that the strip length was chosen correctly.
Transfer impedance $Z_p(f_{RF})=19.0$~Ohm, $Z_p'(f_{RF})=4.15$~kOhm/m.

\section{Beam impedance}
\label{Sec:Beam_impedance}

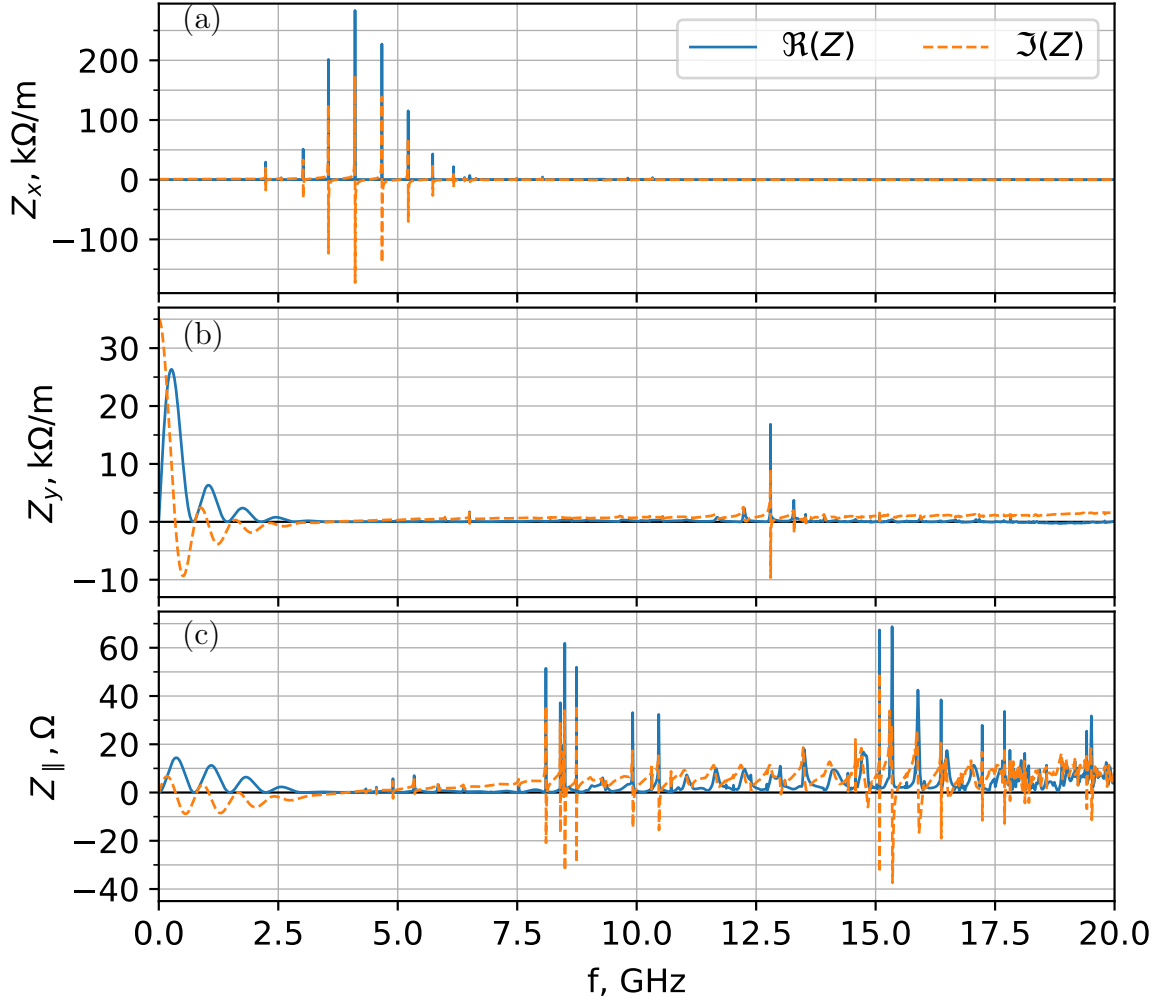
\begin{figure}[h]
    \centering
    \begin{picture}(\textwidth, 0.892\textwidth)(0,0)
    \put(0,0){
    \includegraphics[width=1\textwidth]{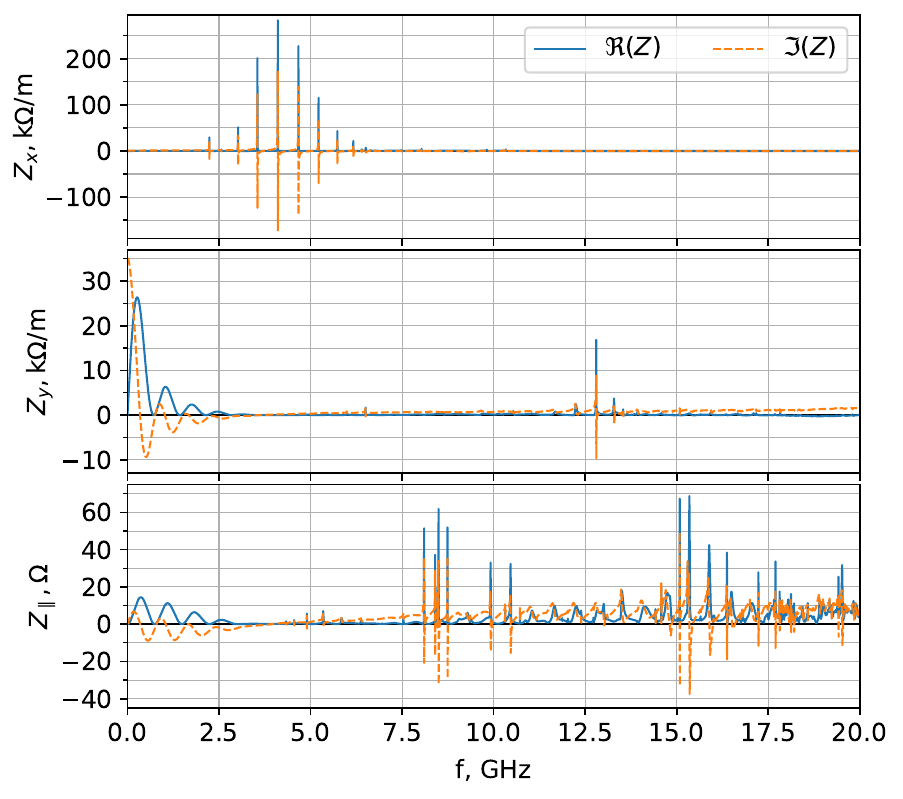}}
    \put(0.17\textwidth, 0.84\textwidth){\makebox(0,0)[lb]{(a)}}
    \put(0.17\textwidth, 0.57\textwidth){\makebox(0,0)[lb]{(b)}}
    \put(0.17\textwidth, 0.313\textwidth){\makebox(0,0)[lb]{(c)}}
    \end{picture}
    \caption{Horizontal (a), vertical (b) and longitudinal (c) impedance of the vertical stripline kicker model.}
    \label{fig:kicker_imp}
\end{figure}

In modern synchrotron radiation sources, collective effects play a significant role, therefore it is important to reduce the impedance of the vacuum chamber elements during the design phase.
For this reason, we chose a stripline kicker design with conically tapered strips at the ends.
With a straight conical taper, both the longitudinal and transverse impedances of the stripline kicker are approximately $\mathrm{sinc}^2(\omega l_t/c_0)$ times smaller than in the case of a straight stripline without tapers~\cite{kicker_M4_v2}.
Here $l_t$ is the length of the conical narrowing.
In our case, the approximation of a straight conical taper is patrly applicable.
The impedance also rapidly decreases to low values in a similar manner.

We calculated  impedance for the vertical stripline kicker model in CST Studio using Wakefield Solver.
Horizontal, vertical, and longitudinal impedance were obtained.
The figure~\ref{fig:kicker_imp} shows the graphs of these impedances.
It is clear that the resulting impedances are small, with the exception of relatively large resonant modes of the horizontal impedance.
This means that the kicker impedance will make the largest contribution to transverse multi-bunch instabilities.

\section{Conclusion}
We developed a stripline kicker for the transverse instability suppression system of SRF ``SKIF'' storage ring.
The stripline kicker usage also as a beam position monitor determined the stripline length.
To reduce the beam impedance of the stripline kicker, we used  design with conically tapered strip ends.
Since the electromagnetic fields of the tapered ends of the stripline are non-negligible, we optimized the stripline kicker for the uniformity of the transverse momentum imparted to the beam particles.
The shape of the striplines was also selected for better matching of the strips with the vacuum feedthroughs in the odd operating mode, when the signals to the kicker strips are supplied with the opposite sign.

Numerical calculations of all key characteristics were performed for the stripline kicker model.
The transverse momentum imparted to the beam particles, as well as the transverse shunt impedance, were calculated for two extreme operating modes: stationary and maximum frequency.
The longitudinal and transverse transfer impedances were calculated as characteristics of the stripline BPM.
Also, for the stripline kicker model, simulations were performed in CST Studio using Wakefield Solver, and the longitudinal and transverse impedances were calculated.

\section{Acknowledgments}
This work was partially supported by the Ministry of Science and Higher Education of the Russian Federation within the governmental order for SRF SKIF Boreskov Institute of Catalysis (project FWUR-2025-0004).

\end{document}